\documentclass[letterpaper]{article} 
\usepackage{aaai25}  
\usepackage{times}  
\usepackage{helvet}  
\usepackage{courier}  
\usepackage[hyphens]{url}  
\usepackage{graphicx} 
\usepackage{natbib}  
\usepackage{caption} 
\usepackage{algorithm}
\usepackage{algorithmic}
\usepackage{graphicx} 
\usepackage{float} 
\usepackage{subfigure} 
\usepackage{diagbox}
\usepackage{algorithm}
\usepackage{algorithmic}
\usepackage{multirow}
\usepackage{makecell}
\usepackage{amsmath}
\usepackage{caption}
\usepackage{cite}
\usepackage{color}
\usepackage{url,array}
\usepackage{amssymb}
\usepackage{booktabs}

\usepackage{newfloat}
\usepackage{listings}
\DeclareCaptionStyle{ruled}{labelfont=normalfont,labelsep=colon,strut=off} 
\floatstyle{ruled}
\newfloat{listing}{tb}{lst}{}
\floatname{listing}{Listing}
\title{StableVC: Style Controllable Zero-Shot Voice Conversion with \\ Conditional Flow Matching}
\author{
    Jixun Yao\textsuperscript{\rm 1}, 
    Yuguang Yang\textsuperscript{\rm 2}, 
    Yu Pan\textsuperscript{\rm 2}, 
    Ziqian Ning\textsuperscript{\rm 1}, 
    Jiaohao Ye\textsuperscript{\rm 2}, 
    Hongbin Zhou\textsuperscript{\rm 2}, 
    Lei Xie\textsuperscript{\rm 1}\thanks{Corresponding author}
}
\affiliations{
    \textsuperscript{\rm 1}
    Audio, Speech and Language Processing Group (ASLP@NPU) \\
School of Computer Science, Northwestern Polytechnical University, Xi'an, China \\
\textsuperscript{\rm 2}Ximalaya Inc, China\\

    \{yaojx,ningziqian,lxie\}@mail.nwpu.edu.cn, \{yuguang.yang,yu.pan,jianhao.ye,hongbin.zhou\}@ximalaya.com
}

\usepackage{bibentry}

\begin{document}

\maketitle

\begin{abstract}
Zero-shot voice conversion (VC) aims to transfer the timbre from the source speaker to an arbitrary unseen speaker while preserving the original linguistic content. 
Despite recent advancements in zero-shot VC using language model-based or diffusion-based approaches, several challenges remain: 1) current approaches primarily focus on adapting timbre from unseen speakers and are unable to transfer style and timbre to different unseen speakers independently; 2) these approaches often suffer from slower inference speeds due to the autoregressive modeling methods or the need for numerous sampling steps; 3) the quality and similarity of the converted samples are still not fully satisfactory.
To address these challenges, we propose a \textbf{St}yle controll\textbf{able} zero-shot VC approach named \textit{StableVC}, which aims to transfer timbre and style from source speech to different unseen target speakers. Specifically, we decompose speech into linguistic content, timbre, and style, and then employ a conditional flow matching module to reconstruct the high-quality mel-spectrogram based on these decomposed features. To effectively capture timbre and style in a zero-shot manner, we introduce a novel dual attention mechanism with an adaptive gate, rather than using conventional feature concatenation. With this non-autoregressive design, StableVC can efficiently capture the intricate timbre and style from different unseen speakers and generate high-quality speech significantly faster than real-time. Experiments demonstrate that our proposed StableVC outperforms state-of-the-art baseline systems in zero-shot VC and achieves flexible control over timbre and style from different unseen speakers. Moreover, StableVC offers approximately 25$\times$ and 1.65$\times$ faster sampling compared to autoregressive and diffusion-based baselines.
\end{abstract}

%

\section{Introduction}

\begin{figure}[ht]
  \centering
  \includegraphics[width=8cm]{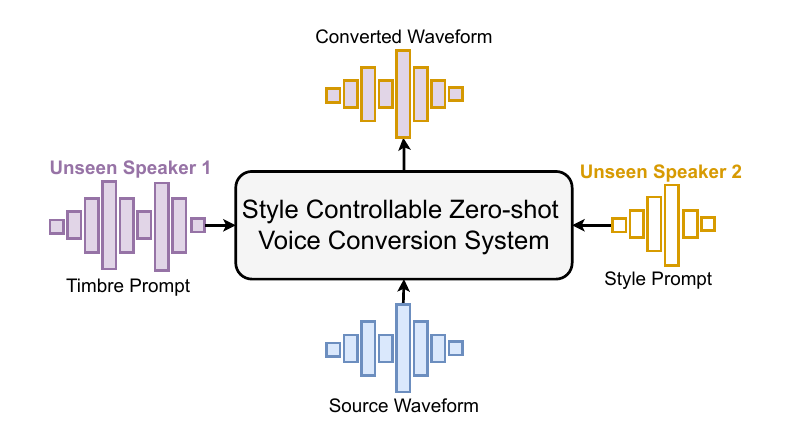}
  \caption{The concept of style-controllable zero-shot voice conversion. It aims to build a VC system capable of adapting timbre to unseen speakers and transferring the style to another unseen speaker. Here, ``unseen" refers to speakers not present in the training set.}
  \label{fig:intro}
  \vspace{-15pt}
\end{figure}

Zero-shot voice conversion (VC) aims to transfer the timbre from the source speaker to an arbitrary unseen speaker while keeping the linguistic content unchanged. Currently, numerous impressive zero-shot VC techniques demonstrate remarkable efficacy in converting realistic and natural sound samples, finding crucial applications in both professional audiobook production and entertainment short videos~\citep{diffvc,ning2023expressive,yao2023preserving,sefvc, choi2024dddm}. 
However, these approaches primarily focus on adapting timbre to unseen speakers while often overlooking style attributes. Style refers to how a person expresses themselves, including pitch variation, speaking rate, intonation, and emotion. It can vary significantly across different contexts to enhance speech expressiveness.
Additionally, most current approaches come at a cost, such as more complicated training setups and computationally expensive autoregressive formulations.

Inspired by the recent successes in the zero-shot capabilities of large-scale language models like GPT~\cite{gpt2}, a similar approach is widely believed to achieve comparable results in zero-shot speech generation. Notably, recent works such as LM-VC~\cite{lmvc} can synthesize high-quality personalized speech with a 3-second enrolled recording from an unseen speaker. However, the style information (such as speaking rate, tone, etc.) is also derived from the 3-second prompt and cannot be independently controlled. An ideal approach would allow for the simultaneous conversion of both timbre and style from the source speech to various arbitrary unseen speakers, enabling free combination, as illustrated in Figure~\ref{fig:intro}. Although some works~\cite{yao2024promptvc,li2023styletts,yuan2021improving} have investigated stylistic voice conversion, most focus on many-to-many scenarios or directly serve timbre as a particular style. Converting both style and timbre to different unseen speakers simultaneously remains a significant challenge in the field.


Inference speed is another challenge of current zero-shot VC approaches. Although conventional VC frameworks can achieve faster speed~\cite{qian2019autovc,wang2021vqmivc,ning2023dualvc,casanova2022yourtts}, the quality and similarity in zero-shot scenarios are unsatisfactory. Recent typical works in zero-shot VC can be divided into two categories: large-language model-based~\cite{valle,lmvc} and diffusion-based approaches~\cite{diffvc,choi2024dddm}. The large-language model-based approach employs a neural audio codec to encode continuous speech waveforms into discrete tokens and training models using a "textless NLP" approach~\cite{borsos2023audiolm}. It demonstrates an impressive ability for in-context learning and performs well in zero-shot speech generation tasks. However, autoregressive token prediction inherently has a slow inference speed, and the discrete token only contains compressed information of the original waveform, hindering its performance on tasks that require high quality and expression~\cite{lee2024dittotts}. On the other hand, the diffusion-based approach is also present in many state-of-the-art speech and audio generation models~\cite{naturalspeech2, liu2023audioldm} but requires multiple reverse steps during inference, which also be computationally intensive.

To achieve style-controllable VC and tackle inference speed problems, we propose StableVC, a style-controllable zero-shot VC model that is fast and efficient. StableVC first disentangles speech into content, timbre, and style, and then employs a flow matching generative module to reconstruct high-quality mel-spectrograms. The flow matching module consists of multiple diffusion transformer (DiT) blocks and is trained to predict a vector field, which efficiently models the probabilistic distribution of the target mel-spectrograms. 
For attribute disentanglement, we use a pre-trained self-supervised model with K-means clustering to extract linguistic content, while a factorized codec is used to extract style representation. 
Instead of using conventional concatenation for timbre and style modeling, we propose a novel dual attention mechanism with an adaptive gate control to capture timbre and style information effectively. An adaptive style gate and strong timbre prior information are introduced to ensure the stability of timbre and style modeling. Therefore, we can convert timbre to the target unseen speaker while flexibly controlling style using another speaker's reference. Audio samples can be found in demo pages~\footnote{\url{https://yaoxunji.github.io/stablevc/}}.
The main contributions of this study can be summarized:
\begin{itemize}
    \item We propose StableVC, a novel style-controllable zero-shot voice conversion approach. To the authors' knowledge, this is the first approach that can independently convert the timbre and style to different unseen speakers, enabling any combination of timbre and style transfer.
    \item We introduce a conditional flow matching module for probability-density path prediction conditioned on the timbre and style, significantly improving synthesis speed and sample quality compared to conventional diffusion-based and language model-based approaches. 
    \item We propose a dual attention mechanism with adaptive gate control, called DualAGC, in the flow matching module to capture distinct timbre and style. An adaptive style gate and timbre prior information are incorporated to ensure the stability of timbre and style modeling. 
\end{itemize}

\section{Related Work}

\subsection{Zero-shot Voice Conversion}
A popular paradigm for zero-shot VC involves tokenizing speech waveforms into discrete semantic and acoustic tokens using a self-supervised learning (SSL) model and a neural audio codec, respectively. This approach allows the generation task to leverage powerful large language models, exhibiting impressive zero-shot capabilities. For example, LM-VC~\cite{lmvc} employs a two-stage language modeling approach: first generating coarse acoustic tokens to recover the source linguistic content and target speaker’s timbre, and then reconstructing fine acoustic details as converted speech. Similarly, several zero-shot VC frameworks~\cite{yang2024takinvc,sefvc,acevc,wang2022drvc} use semantic representations extracted from SSL models to disentangle the linguistic content. GR0~\cite{gr0vc} uses a generative SSL framework to jointly learn a global speaker embedding and a zero-shot voice converter, achieving zero-shot VC. Another approach~\cite{timbrevc1} decouples speech into local and global representations and employs dropout with multiplicative Gaussian noise to apply an information bottleneck for timbre disentanglement. Although these approaches achieve zero-shot VC, they mainly focus on capturing the timbre and overlook the style information.

Several works have dedicated significant efforts to exploring the simultaneous modeling of fundamental frequency (F0) and timbre~\cite{choi2024dddm}. For example, DVQVC~\cite{dvqvc} and SLMGAN~\cite{li2023slmgan} concatenate F0 with linguistic content and speaker timbre to reconstruct the target speech. Diff-HierVC~\cite{choi2023diff} introduces a hierarchical system to generate F0 with the target voice style and convert the speech based on the generated F0. Additionally, VoiceShopVC~\cite{voiceshopvc} employs a conditional diffusion backbone model with optional normalizing flow-based modules to achieve speaker attribute editing. However, these works primarily focus on F0 not style and scenarios are many-to-many, which can not adapt timbre and style from different unseen speakers.

\begin{figure*}[ht]
  \centering
  \includegraphics[width=13
  cm]{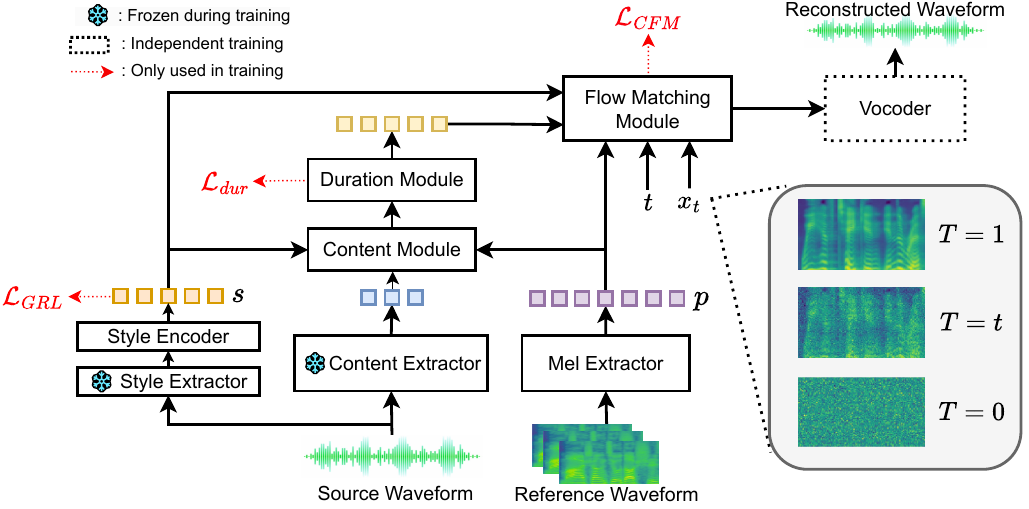}
  \caption{The overall framework of StableVC includes three feature extractors for style, linguistic content, and mel-spectrogram extraction. It also incorporates a content module and a duration module to re-predict the duration based on different styles and timbres, and a flow matching module generates high-quality speech at speeds significantly faster than real-time.}
  \label{fig:model}
  \vspace{-15pt}
\end{figure*}

\subsection{Flow Matching}
Flow matching generative models~\cite{flowmatching} estimate the vector field of the transport probability path from noise to the target distribution. These models learn the transport path using an ordinary differential equation (ODE) to find a straight path that connects the noise and target samples, thereby reducing transport costs and requiring fewer sampling steps. Compared to conventional denoising diffusion models like DDPM~\cite{ddpm}, flow matching provides more stable training and superior inference speed. This technique has shown excellent performance in accelerating image generation~\cite{stablediffusion3}.

In speech generation, Voicebox~\cite{le2024voicebox} leverages flow matching to build a large-scale, text-conditioned speech generative model. Its successor, Matcha-TTS~\cite{matchatts}, adopts an encoder-decoder architecture and utilizes optimal transport flow matching for model training. VoiceFlow~\cite{guo2024voiceflow} treats the generation of mel-spectrograms as an ordinary differential equation conditioned on text inputs, incorporating rectified flow matching to boost sampling efficiency. Despite these advancements, flow matching has not yet been explored in zero-shot VC to enhance converted quality and inference speed.

\section{StableVC}
\subsection{Overview}
We aim to provide flexible style control capabilities in zero-shot VC with high quality and efficiency.
The overall framework of StableVC is illustrated in Figure~\ref{fig:model}. First, we extract linguistic content, style representation, and mel-spectrogram from the waveform. To disentangle linguistic content, we use a pre-trained WavLM model\footnote{\url{https://github.com/microsoft/unilm/tree/master/wavlm}} and apply K-means clustering to extract discrete tokens~\cite{chen2022wavlm}, setting the number of K-means clusters to 1024. Additionally, we deduplicate adjacent tokens and replace them with the corresponding K-means embeddings. This process is the same as described in \citealp{yao2024promptvc} and has demonstrated its effectiveness in linguistic content extraction.
The purpose of deduplication is to re-predict the duration for each token, allowing the duration of the same linguistic content to adapt to different timbre and styles. For style representation, we use a factorized codec\footnote{\url{https://github.com/open-mmlab/Amphion/tree/main/models/codec/ns3_codec}} as style extractor. This model can extract disentangled subspace representations of speech style. For timbre representation, we extract the mel-spectrogram from multiple reference speeches by the same speaker as the source speech. The details of timbre and style modeling will be discussed in the following section.

The disentangled features are encoded through the content module, which consists of multiple DiT blocks. This is followed by the duration module, which predicts the duration and aligns the hidden representations to the frame level. The output from the duration module is transformed into mel-spectrograms using the flow matching module, which comprises multiple DiT blocks with timestep fusion. Style and timbre are modeled by the DualAGC within each DiT block. Finally, the generated mel-spectrograms are reconstructed into waveforms using an independently trained vocoder.

\subsection{DualAGC}
\begin{figure}[ht]
  \centering
  \includegraphics[width=7cm]{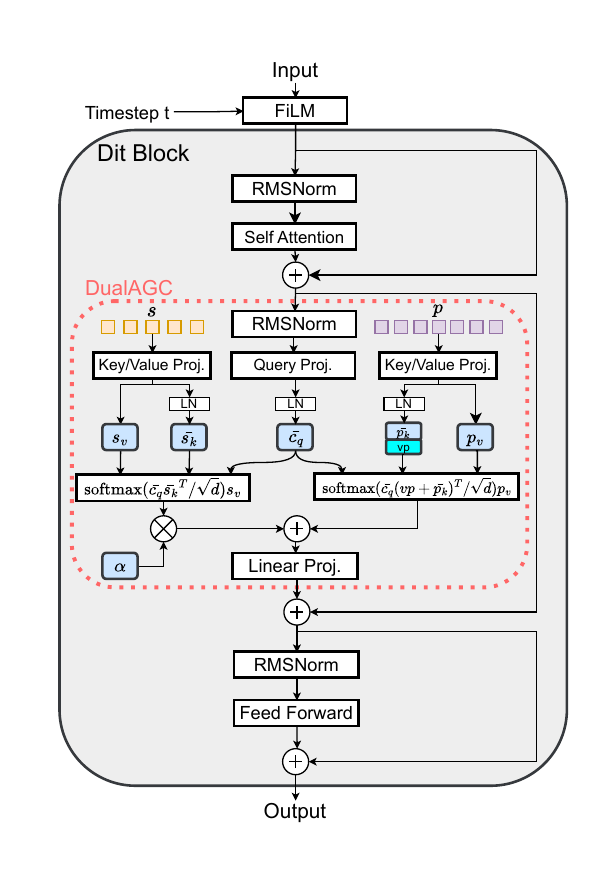}
  \caption{Details of DualAGC in the DiT block.}
  \label{fig:dit}
  \vspace{-10pt}
\end{figure}

The widely used DiT block employs adaptive layer norm or cross-attention to append external conditions, typically including only timesteps and one instructive condition~\cite{dit}. In this section, we introduce DualAGC, which incorporates a dual attention approach to capture style information and speaker timbre simultaneously. To improve the stability of timbre modeling, we introduce timbre prior information in the dual attention mechanism. Additionally, an adaptive gate mechanism is employed to gradually integrate style information into the content and timbre.
The detail of DualAGC in the DiT block is shown in Figure~\ref{fig:dit}.

Suppose the input of the DiT block is $c$ and the outputs of the style encoder and mel extractor are $s$ and $p$, respectively. We add a FiLM layer~\cite{perez2018film} before the DiT block to apply an affine feature-wise transformation conditioned on timestep $t$. The quantized style embeddings extracted from the style extractor are passed through the style encoder, which consists of an average pooling layer and several convolutional blocks. The style encoder compresses the quantized embedding four times in the time dimension and captures the correlation of style information~\cite{jiang2024megatts2}. $p$ is the mel-spectrogram extracted from multiple reference speech, while $\bar{c_q}$ represents the hidden representation obtained by query projection and query-key normalization~\cite{qknorm}.

\textbf{Timbre Attention}: 
Our objective is to extract fine-grained information from reference speech that represents the speaker's timbre. We employ $\bar{c_q}$ as the attention query and use the mel-spectrogram extracted from multiple references as the value of timbre attention. The cross-attention mechanism is agnostic to input positions, effectively resembling a temporal shuffling of the mel-spectrogram. This process preserves a significant amount of speaker information while minimizing other details, such as linguistic content, enabling the cross-attention mechanism to focus on learning and capturing speaker timbre from the reference speech~\cite{sefvc}. To improve the stability of timbre modeling, we extract a global speaker embedding from a pre-trained speaker verification model
and concatenate it with the mel-spectrogram as the attention key. The extracted speaker embedding serves as an instructive timbre prior, guiding the timbre attention to capture timbre-related information. 
\vspace{-10pt}

\textbf{Style Attention}: The compressed style representations $s$ serve as both the key and value in style attention, with $\bar{c_q}$ as the attention query. To gradually inject style information into linguistic content and timbre, we propose an adaptive gating mechanism and employ a zero-initialized learnable parameter $\alpha$ as the gate to control the injection process. 
Given the style keys $s_k$ and values $s_v$ with timbre keys $p_k$ and values $p_v$, the final output $O$ of DualAGC is formulated as:
\begin{equation}
    O = \text{softmax}(\frac{\bar{c_q}(\text{vp}\bigoplus\bar{p_k})^T}{\sqrt{d}})p_v + \text{tanh}(\alpha)\text{softmax}(\frac{\bar{c_q}\bar{s_k}^T}{\sqrt{d}})s_v,
\end{equation}
where $\bar{s_k}$ and $\bar{p_k}$ stand for applying query-key norm, $d$ is the dimension of queries and keys. $\bigoplus$ and $\text{vp}$ represent concatenation and the global speaker embedding extracted from the pre-trained speaker verification model. The style injection process allows for adaptive style modeling without impacting the final performance of timbre modeling. 

\subsection{Conditional Flow Matching}
In this section, we first introduce the probability-density path generated by a vector field and then lead into the training objective used in our flow matching module. 

Flow matching~\cite{flowmatching,conditional_flowmatching} is a method used to fit the time-dependent probability path between the target distribution $p_1(x)$ and a standard distribution $p_0(x)$. It is closely related to continuous normalizing flows but is trained more efficiently in a simulation-free fashion. 
The flow $\phi:[0,1] \times \mathbb{R}^d \rightarrow \mathbb{R}^d$ is defined as the mapping between two density functions using the ODE:
\begin{equation}
    \frac{d}{d t} \phi_t(x)=v_t\left(\phi_t(x)\right) ; \quad \phi_0(x)=x
\end{equation}
where $v_t(x)$ represents the time-dependent vector field and is also a learnable component. To efficiently sample the target distribution in fewer steps, we employ conditional flow matching with optimal transport as specified in~\citealp{conditional_flowmatching}. 
Since it is difficult to determine the marginal flow in practice, we formulate it by marginalizing over multiple conditional flows as follows:
\begin{equation}
    \phi_{t, x_1}(x)=\sigma_t\left(x_1\right) x+\mu_t\left(x_1\right),
\end{equation}
where $\sigma_t(x)$ and $\mu_t(x)$ are time-conditional affine transformations used to parameterize the transformation between distributions $p_1(x)$ and $p_0(x)$. 
For the unknown distribution $q(x)$ over our training data, we define $p_1(x)$ as the approximation of $q(x)$ by perturbing individual samples with small amounts of white noise with $\sigma_{\text{min}}=0.0001$.

Therefore, we can specify our trajectories with simple linear trajectories as follows:
\begin{equation}
    \mu_t(x)=t x_1, \sigma_t(x)=1-\left(1-\sigma_{\min }\right) t.
\end{equation}
The final training objective of the vector field using the conditional flow matching is denoted as :
\begin{small}
\begin{equation}
    \mathcal{L}_{\text{CFM}}=\mathbb{E}_{t, q\left(x_1\right), p\left(x_0\right)}\left\|v_t\left(\phi_{t, x_1}\left(x_0\right) ; h\right)-\left(x_1-\left(1-\sigma_{\min }\right) x_0\right)\right\|^2
\end{equation}
\end{small}
where $h$ is the conditional set containing the output of the duration module and the style and timbre representations.

Conditional flow matching encourages simpler and straighter trajectories between source and target distributions without the need for additional distillation. We sample from standard Gaussian distribution $p_0(x)$ as the initial condition at $t = 0$. Using 10 Euler steps, we approximate the solution to the ODE and efficiently generate samples that match the target distribution.

\subsection{Training Objectives}
In practice, even though the style representation extracted from the factored codec primarily contains style-related information, there may still be potential timbre leakage. To mitigate this risk and achieve better disentanglement between style and timbre, we employ an adversarial classifier with a gradient reversal layer (GRL)~\cite{GRL} to eliminate potential timbre information in the output of the style encoder. We first average the output of the style encoder into a fixed-dimensional global representation and employ an additional speaker classifier to predict its speaker identity. The GRL loss can be denoted as follows:
\begin{equation}
    \mathcal{L}_{\text{GRL}}=\mathbb{E}[-log(C_\theta(I\mid \text{avg}(s)))], 
\end{equation}
where $C_\theta$ and $I$ represent the speaker classifier and speaker identity label. The gradients are reversed to optimize the style encoder to eliminate potential timbre information.

The duration module is trained to re-predict the duration (deduplicated length in the content extractor) conditioned on the style and timbre representations. This enables the generated waveform to adapt different durations based on the different timbre and style. The training objective $\mathcal{L}_{\text{dur}}$ for the duration module is to minimize the mean squared error with respect to the log-scale predicted duration and the ground-truth duration.
The overall training loss of StableVC is:
\begin{equation}
    \mathcal{L}= \mathcal{L}_{\text{CFM}}+\mathcal{L}_{\text{dur}}+\lambda\mathcal{L}_{\text{GRL}},
\end{equation}
where $\lambda$ is the hyper-parameter used to balance the loss term and we set $\lambda=0.1$.

\section{Experimental Setup}
\subsection{Configuration}
We train StableVC on 8 NVIDIA 3090 GPUs for 800K iterations with a total batch size of 128 and the AdamW optimizer is used with a learning rate of 0.0001. 
During inference, we sample the target mel-spectrograms using 10 Euler steps in the flow matching module with a guidance scale of 1. The mel-spectrograms are reconstructed to waveform by Hifi-GAN vocoder~\cite{hifigan}.

\subsection{Datsets}
We conduct our experiments on the LibriLight dataset~\cite{librilight}, which consists of 60k hours of speech data. For model training, we use samples longer than 5 seconds and filter out low-quality samples using DNSMOS P.808 scores~\cite{reddy2022dnsmos}, resulting in a 20k hours subset. For evaluation, we use the VCTK corpus~\cite{vctk} and ESD corpus~\cite{zhou2022emotional}, ensuring no speaker or style overlap with the training data. 
The timbre references are selected from the VCTK dataset, consisting of 10 male and 10 female speakers. The style references are selected from the ESD dataset and we chose two high-intensity samples for each of five different styles, resulting in 200 reference samples.

\subsection{Baseline Systems}
We conduct a comparative analysis of the performance in zero-shot voice conversion between our proposed StableVC approach and several baseline systems, encompassing the following system: 
1) StyleVC~\cite{stylevc}: a style voice conversion system that employs adversarial style generalization;
2) LM-VC~\cite{lmvc}: a two-stage language model based approach for zero-shot voice conversion; 
3) VALLE-VC~\cite{valle}, a language model based speech synthesis approach where we replace the original phoneme input with the same content representation in StableVC to enable voice conversion capabilities; 
4) NS2VC\footnote{\url{https://github.com/adelacvg/NS2VC}}, a voice conversion version of NaturalSpeech2~\cite{naturalspeech2}; 
5) DDDM-VC~\cite{choi2024dddm}, a high-quality zero-shot voice conversion system based on decoupled denoising diffusion models;
6) SEF-VC~\cite{sefvc}, a speaker embedding free zero-shot voice conversion model.

\subsection{Evaluation Metrics}
\textbf{Subjective Metrics}: 
We employ the naturalness mean opinion score (nMOS) to evaluate the naturalness of the generated samples. Additionally, we use two similarity mean opinion scores (sMOS-s and sMOS-p) to evaluate style similarity and timbre similarity, respectively. 

\begin{table*}[ht]
\centering
\caption{The subjective and objective evaluation results for StableVC and the baseline systems in zero-shot voice conversion. All subjective metrics are computed with 95\% confidence intervals and ``GT" refers to ground truth samples.}
\label{tab:timbre}
\vspace{-5pt}
\renewcommand\arraystretch{1.2}
\resizebox{0.75\linewidth}{!}{
\begin{tabular}{lccccccc}
\hline
         & nMOS $\uparrow$ & sMOS-p $\uparrow$      & UTMOS $\uparrow$        & WER $\downarrow$ & SECS $\uparrow$ & RTF $\downarrow$ & \#Param. \\ \hline
GT       & 4.33$\pm0.04$ &   -        & 4.24          & 1.61 & -    & -              & -        \\ \hline
LMVC     & 3.01$\pm0.05$ & 3.17$\pm0.07$       & 3.32          & 4.17 & 0.61 & 3.891          & 305M     \\
VALLE-VC & 3.56$\pm0.05$ & 3.65$\pm0.04$       & 3.63          & 3.08 & 0.55 & 3.944          & 302M     \\
StyleVC   & 3.24$\pm0.07$ & 3.21$\pm0.07$       & 3.41          & 5.21 & 0.43 & \textbf{0.075}          & 31M     \\
NS2VC    & 3.32$\pm0.05$ & 3.16$\pm0.05$       & 3.49          & 4.88 & 0.44 & 0.337          & 435M     \\
DDDM-VC  & 3.67$\pm0.07$ & 3.61$\pm0.06$       & 3.75          & 3.07 & 0.51 & 0.287          & 66M      \\
SEF-VC   & 3.63$\pm0.04$ & 3.72$\pm0.05$       & 3.59          & 2.89 & 0.53 & 0.168          & 260M     \\ \hline
StableVC & \textbf{3.96$\pm$0.04} & \textbf{4.04$\pm$0.05}       & \textbf{4.12}          & \textbf{2.03} & \textbf{0.67} & 0.146          & 166M     \\ \hline
\end{tabular}
}
\vspace{-15pt}
\end{table*}

\textbf{Objective Metrics}: For objective evaluation, we employ cosine distance (SECS) for speaker similarity, word error rate (WER) for robustness and MOS predicted by neural network (UTMOS) for speech quality. In specific, 1) SECS: we employ the WavLM-TDCNN speaker verification model\footnote{\url{https://github.com/microsoft/UniSpeech/tree/main/downstreams/speaker_verification}} to evaluate speaker similarity between generated samples and the target speaker reference; 2) WER: we use a pre-trained CTC-based ASR model\footnote{\url{https://huggingface.co/facebook/hubert-large-ls960-ft}} to transcribe the generated speech and compare with ground-truth transcription; 3) UTMOS~\cite{utmos}: a MOS prediction system that ranked first in the VoiceMOS Challenge 2022\footnote{\url{https://github.com/tarepan/SpeechMOS}}. It is used to estimate the speech quality of the generated samples. 

For inference latency, we calculate the real-time factor (RTF) on a single NVIDIA 3090 GPU to compare latency across systems. Additionally, we compute the number of model parameters (\#Param.) for reference. Since all baseline systems use SSL models to extract representations and employ a vocoder or codec to reconstruct the waveform, we only count the parameters of the acoustic model.
We employ two pitch-related metrics for the evaluation of style transfer: Root Mean Squared Error (RMSE) and Pearson correlation (Corr)~\cite{sedgwick2012pearson}. These metrics are widely used in the evaluation of stylistic VC. Since the sequences between the reference and the generated samples are not aligned, we perform Dynamic Time Warping~\cite{muller2007dynamic} to align the sequences before comparison.

\section{Experimental Results}
\subsection{Experimental Results on Zero-shot VC}
In this subsection, we first evaluate the performance in zero-shot voice conversion and compare StableVC with baselines in terms of speech naturalness, quality, speaker similarity, robustness, and inference latency. The evaluation results of both subjective and objective metrics are shown in Table~\ref{tab:timbre}.

For the naturalness of the converted samples, StableVC outperforms the baseline systems by a substantial margin, achieving nMOS scores closest to the ground truth samples. Regarding speech quality, StableVC attains the highest UTMOS score of 4.12, surpassing all baseline systems and showing only a slight decline compared to the ground truth. These results demonstrate that samples converted by our proposed StableVC are both natural and high-quality.

The speaker similarity results show that: 1) StableVC achieves an sMOS-p score of 4.05 and an SECS of 0.67, reflecting the high speaker similarity of the generated sample; 2) most baseline systems only obtain around 0.5 of SECS, which is significantly lower than StableVC, while the sMOS-p for baseline systems also reflects a similar trend. These findings highlight the superiority of our approach in timbre modeling and also confirm that using attention mechanisms to capture fine-grained timbre details leads to a substantial enhancement in speaker similarity.

We evaluate the robustness of StableVC in zero-shot VC by computing the WER of converted speech samples, as shown in Table~\ref{tab:timbre}. The results show that: 1) StableVC achieves a 2.03 WER, only about 0.4 higher than the ground truth samples, proving its high robustness and intelligibility; 2) StableVC outperforms other baselines by a considerable margin, demonstrating the superior robustness of StableVC.

\vspace{-5pt}
\begin{table}[ht]
\centering
\caption{Objective metrics comparison of zero-shot style transfer between StableVC and baseline systems.}
\label{tab:style}
\renewcommand\arraystretch{1.2}
\resizebox{1.0\linewidth}{!}{
\begin{tabular}{lccccc}
\hline
Model    & Corr $\uparrow$   & RMSE $\downarrow$  & UTMOS $\uparrow$ & WER $\downarrow$      & SECS $\uparrow$         \\ \hline
StyleVC  & 0.71   & 14.98  & 3.63   & 3.54 & 0.21          \\
NS2VC    & 0.67   & 17.19  & 3.57   & 5.31 & 0.47          \\
DDDM-VC    & 0.69   & 15.68  & 3.71   & 3.56 & 0.50          \\
StableVC & \textbf{0.75}   & \textbf{12.87}  & \textbf{4.06}   & \textbf{2.12} & \textbf{0.64} \\ \hline
\end{tabular}}
\vspace{-10pt}
\end{table}

To compare inference latency, we measure the RTF against various baseline systems and count the total trainable parameters of each system for reference. The results show that StableVC achieves approximately a 25.3$\times$ speedup over the language model-based approach and about a 1.65$\times$ speedup over the denoising diffusion-based approach, while consistently surpassing these baseline systems in all subjective and objective metrics. This demonstrates that StableVC is both effective and efficient. Although StyleVC has the lowest RTF and smallest model parameters, its performance on each metric is not ideal. Moreover, unlike the autoregressive baseline, where generation time is proportional to the source speech duration, the non-autoregressive StableVC maintains similar latency regardless of the length.
These results highlight the efficiency of StableVC in zero-shot voice conversion, offering a significant advantage in terms of both speed and performance.

\vspace{-5pt}
\subsection{Experimental Results on Style Transfer}
In this subsection, we evaluate the performance of zero-shot style transfer, where the timbre and style prompts are from different unseen speakers. Since some baseline systems do not support both timbre and style prompts, we select StyleVC, NS2VC, and DDDM-VC as baseline systems. For these systems, we use the F0 extracted from the style reference as the style prompt and compare these baselines with StableVC using subjective and objective metrics.

\vspace{-5pt}
\begin{figure}[ht]
  \centering
  \includegraphics[width=8cm]{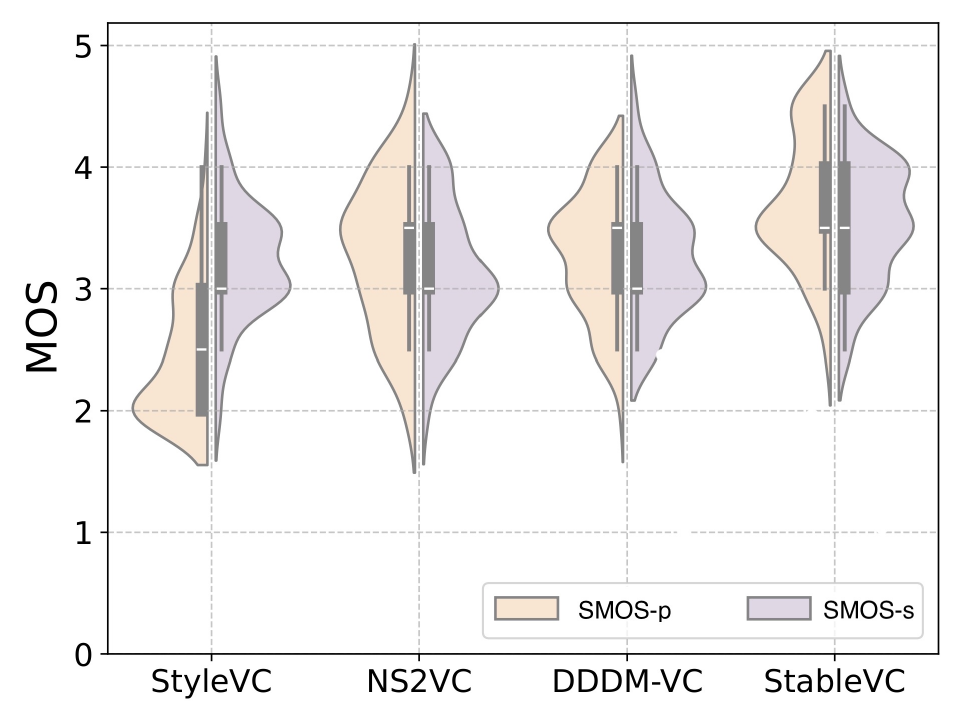}
  \caption{Violin plots for timbre and style similarity of speech generated by baseline systems and StableVC.}
  \label{fig:pinao}
  \vspace{-10pt}
\end{figure}

Table~\ref{tab:style} shows the objective metrics comparison between  StableVC and baseline systems. We have the following observations: 1) StableVC outperforms baseline systems across all measured metrics, including pitch-related metrics, speech quality, and speaker similarity; 2) StyleVC achieves comparable results in style transfer but struggles with effectively transferring the target timbre. The lower speaker timbre similarity in the converted samples of StyleVC is primarily due to timbre leakage during the style modeling process. This leakage causes the style features to carry some timbre information, thereby compromising the accuracy of the timbre conversion.

Furthermore, we visualize the subjective evaluation results of timbre similarity and style similarity using violin plots, as shown in Figure~\ref{fig:pinao}. These plots indicate that the median values for both style similarity and speaker similarity from StableVC are higher than those of the three baseline systems. Additionally, the overall distribution of subjective scores for our proposed StableVC is superior to the baseline systems. Notably, the speaker similarity scores for StyleVC are significantly lower in subjective evaluations.

The subjective evaluation results are consistent with the objective evaluation results. These results demonstrate the superior performance of StableVC in accurately transferring both timbre and style, achieving higher speaker similarity, lower WER, and more accurate style representation compared to the baseline systems. These metrics highlight StableVC's capability to independently and accurately convert timbre and style from different unseen speakers, leading to high-quality, intelligible, and expressive conversion results.

\begin{table}[ht]
\centering
\caption{Effect of Euler sample steps N on flow matching module. Objective metrics improve rapidly in the first 5 steps and continued qualitative improvements up to 20 steps.}
\label{tab:ablation_step}
\renewcommand\arraystretch{1.1}
\resizebox{0.95\linewidth}{!}{
\begin{tabular}{lccccc}
\hline
Model                     & N & \multicolumn{1}{c}{UTMOS $\uparrow$} & \multicolumn{1}{c}{WER $\downarrow$} & \multicolumn{1}{c}{SECS $\uparrow$} & \multicolumn{1}{c}{RTF $\downarrow$} \\ \hline
\multirow{4}{*}{StableVC} & 1    & 3.42     & 3.87                    & 0.54                     & 0.031                              \\
                          & 2    & 3.49     & 3.94                    & 0.55                     & 0.043                              \\
                          & 5    & 3.82     & 2.98                    & 0.61                     & 0.068                              \\
                          & 10   & 4.12     & 2.03                    & 0.67                     & 0.146                              \\
                          & 20   & 4.15     & 2.07                    & 0.66                     & 0.215                              \\\hline
\end{tabular}}
\vspace{-15pt}
\end{table}

\subsection{Ablation Study}
In this subsection, we conduct ablation studies to verify the following: 1) the effect of using different Euler sampling steps on the quality and similarity of the converted samples; 2) the effectiveness of each method in timbre and style modeling, including the introduction of the timbre prior information and multiple reference speech in timbre attention, the adaptive gate in style attention, the GRL loss, and the deduplication for re-predicting duration. These studies aim to evaluate the contributions of each component to the overall performance.

We compare the performance by using Euler steps ranging from 1 to 20 to evaluate the trade-offs between efficiency and quality, as shown in Table~\ref{tab:ablation_step}. By analyzing these results, we observe that StableVC achieves low WER even with very few Euler steps, indicating that the sample generated by the flow matching module can be accurately recognized by the ASR model, demonstrating high robustness and intelligibility. Evaluation results stabilize after approximately 5 Euler steps, with better audio quality achieved around 10 steps and only slight improvement observed at 20 steps. Although the RTF increases with the number of sampling steps, the RTF at 10 Euler steps remains competitive compared to most baseline systems, as shown in Table~\ref{tab:timbre}.

\begin{table}[ht]
\centering
\caption{The ablation study results of removing each component in timbre and style modeling. 
}
\label{tab:ablation_modeling}
\renewcommand\arraystretch{1.2}
\resizebox{1.0\linewidth}{!}{
\begin{tabular}{lccccc}
\hline
Model         & Corr $\uparrow$ & RMSE $\downarrow$ & UTMOS $\uparrow$ & WER $\downarrow$ & SECS $\uparrow$   \\ \hline
StableVC      & 0.75 & 12.87 & 4.06 & 2.12 & 0.64  \\
\quad w/o $\text{vp} \& \text{multi}$ & 0.59 & 18.78 & 3.51 &22.16 & 0.41 \\
\quad w/o $\alpha$  & 0.73 & 12.85 & 4.01 &2.21 & 0.36  \\ 
\quad w/o $\mathcal{L_{\text{GRL}}}$  & 0.68 & 14.61 & 3.62 &2.63 & 0.51  \\ 
\quad w/o \text{dur}  & 0.66 & 16.51 & 4.05 &2.27 & 0.63  \\ \hline
\end{tabular}}
\vspace{-10pt}
\end{table}


To verify the effectiveness of each method in timbre and style modeling, we conduct the following ablation studies: 1) only use the source utterance as reference at the training stage and do not concatenate with timbre prior vp, denoted as \textbf{w/o $\text{vp} \& \text{multi}$}; 2) remove the adaptive gate in style attention, denoted as \textbf{w/o $\alpha$}; 3) remove the GRL loss, denoted as \textbf{w/o $\mathcal{L_{\text{GRL}}}$}; 4) remove the deduplication in linguistic content extraction and duration modeling, denoted as \textbf{w/o \text{dur}}.

As shown in Table~\ref{tab:ablation_modeling}, we observe significant performance degradation across all metrics when the multiple reference and timbre prior are removed, particularly in the WER results. The primary reason is that the timbre attention mechanism captures irrelevant linguistic information from the reference speech, leading to poor intelligibility in the final converted samples. This degradation highlights the critical role of the multiple reference speech and timbre prior in capturing timbre information.

A noticeable decline in SECS is observed when the adaptive gate is removed. This indicates that without the adaptive gate, the style attention mechanism captures extraneous information from the style prompt, weakening the relationship between the attention score and style information, resulting in lower speaker similarity. These findings suggest that the adaptive gate plays a critical role in controlling style injection without compromising timbre similarity, even when style representations from unseen speakers are injected. When the GRL loss is removed, both style and timbre metrics degrade. This demonstrates that the GRL loss aids in disentangling timbre and style information, thereby enhancing the final timbre and style similarity of the generated samples. Additionally, a decline in style-related metrics is observed when deduplication and duration modeling are removed. By comparing the performance of these ablated models with the full StableVC model, we can evaluate the contribution of each component to the system’s overall ability to control timbre and style information.

\section{Conclusion}
In this study, we propose StableVC to flexibly and efficiently convert style and timbre of source speech to different unseen target speakers. StableVC consists of 1) three feature extractors to disentangle source speech into linguistic content, style, and timbre, and 2) a conditional flow matching module to reconstruct the target mel-spectrogram in a non-autoregressive manner. We propose a novel dual attention mechanism with an adaptive gate in the DiT block to effectively capture distinctive timbre and style information. Experimental results demonstrate that StableVC outperforms all zero-shot VC baseline systems in both subjective and objective metrics. We also show that StableVC enables flexible conversion of style and timbre in a zero-shot manner. For inference efficiency, StableVC offers approximately 25$\times$ and 1.65$\times$ faster inference speed compared to autoregressive and diffusion baselines.

\bibliography{aaai25}

\begin{thebibliography}{50}
\providecommand{\natexlab}[1]{#1}

\bibitem[{Anastassiou et~al.(2024)Anastassiou, Tang, Peng, Jia, Li, Tu, Wang, Wang, and Ma}]{voiceshopvc}
Anastassiou, P.; Tang, Z.; Peng, K.; Jia, D.; Li, J.; Tu, M.; Wang, Y.; Wang, Y.; and Ma, M. 2024.
\newblock VoiceShop: A Unified Speech-to-Speech Framework for Identity-Preserving Zero-Shot Voice Editing.
\newblock \emph{arXiv preprint arXiv:2404.06674}.

\bibitem[{Borsos et~al.(2023)Borsos, Marinier, Vincent, Kharitonov, Pietquin, Sharifi, Roblek, Teboul, Grangier, Tagliasacchi et~al.}]{borsos2023audiolm}
Borsos, Z.; Marinier, R.; Vincent, D.; Kharitonov, E.; Pietquin, O.; Sharifi, M.; Roblek, D.; Teboul, O.; Grangier, D.; Tagliasacchi, M.; et~al. 2023.
\newblock {AudioLM}: a language modeling approach to audio generation.
\newblock \emph{IEEE/ACM Trans. on Audio, Speech, and Lang. Process.}, 31: 2523--2533.

\bibitem[{Casanova et~al.(2022)Casanova, Weber, Shulby, Junior, G{\"o}lge, and Ponti}]{casanova2022yourtts}
Casanova, E.; Weber, J.; Shulby, C.~D.; Junior, A.~C.; G{\"o}lge, E.; and Ponti, M.~A. 2022.
\newblock Yourtts: Towards zero-shot multi-speaker tts and zero-shot voice conversion for everyone.
\newblock In \emph{Proc. ICML}, 2709--2720.

\bibitem[{Chen et~al.(2022)Chen, Wang, Chen, Wu, Liu, Chen, Li, Kanda, Yoshioka, Xiao et~al.}]{chen2022wavlm}
Chen, S.; Wang, C.; Chen, Z.; Wu, Y.; Liu, S.; Chen, Z.; Li, J.; Kanda, N.; Yoshioka, T.; Xiao, X.; et~al. 2022.
\newblock Wavlm: Large-scale self-supervised pre-training for full stack speech processing.
\newblock \emph{IEEE Journal of Selected Topics in Signal Processing}, 16(6): 1505--1518.

\bibitem[{Choi, Lee, and Lee(2023)}]{choi2023diff}
Choi, H.-Y.; Lee, S.-H.; and Lee, S.-W. 2023.
\newblock Diff-HierVC: Diffusion-based hierarchical voice conversion with robust pitch generation and masked prior for zero-shot speaker adaptation.
\newblock \emph{International Speech Communication Association}, 2283--2287.

\bibitem[{Choi, Lee, and Lee(2024)}]{choi2024dddm}
Choi, H.-Y.; Lee, S.-H.; and Lee, S.-W. 2024.
\newblock Dddm-vc: Decoupled denoising diffusion models with disentangled representation and prior mixup for verified robust voice conversion.
\newblock In \emph{Proc. AAAI}, volume~38, 17862--17870.

\bibitem[{Esser et~al.(2024)Esser, Kulal, Blattmann, Entezari, M{\"u}ller, Saini, Levi, Lorenz, Sauer, Boesel et~al.}]{stablediffusion3}
Esser, P.; Kulal, S.; Blattmann, A.; Entezari, R.; M{\"u}ller, J.; Saini, H.; Levi, Y.; Lorenz, D.; Sauer, A.; Boesel, F.; et~al. 2024.
\newblock Scaling rectified flow transformers for high-resolution image synthesis.
\newblock In \emph{Proc. ICML}.

\bibitem[{Ganin and Lempitsky(2015)}]{GRL}
Ganin, Y.; and Lempitsky, V. 2015.
\newblock Unsupervised domain adaptation by backpropagation.
\newblock In \emph{Proc. ICML}, 1180--1189.

\bibitem[{Guo et~al.(2024)Guo, Du, Ma, Chen, and Yu}]{guo2024voiceflow}
Guo, Y.; Du, C.; Ma, Z.; Chen, X.; and Yu, K. 2024.
\newblock Voiceflow: Efficient text-to-speech with rectified flow matching.
\newblock In \emph{Proc. ICASSP}, 11121--11125.

\bibitem[{Henry et~al.(2020)Henry, Dachapally, Pawar, and Chen}]{qknorm}
Henry, A.; Dachapally, P.~R.; Pawar, S.; and Chen, Y. 2020.
\newblock Query-key normalization for transformers.
\newblock \emph{arXiv preprint arXiv:2010.04245}.

\bibitem[{Ho, Jain, and Abbeel(2020)}]{ddpm}
Ho, J.; Jain, A.; and Abbeel, P. 2020.
\newblock Denoising diffusion probabilistic models.
\newblock \emph{Proc. {NIPS}}, 33: 6840--6851.

\bibitem[{Hussain et~al.(2023)Hussain, Neekhara, Huang, Li, and Ginsburg}]{acevc}
Hussain, S.; Neekhara, P.; Huang, J.; Li, J.; and Ginsburg, B. 2023.
\newblock {ACE-VC}: Adaptive and controllable voice conversion using explicitly disentangled self-supervised speech representations.
\newblock In \emph{Proc. {ICASSP}}, 1--5.

\bibitem[{Hwang, Lee, and Lee(2022)}]{stylevc}
Hwang, I.-S.; Lee, S.-H.; and Lee, S.-W. 2022.
\newblock StyleVC: Non-Parallel Voice Conversion with Adversarial Style Generalization.
\newblock In \emph{Proc. ICPR}, 23--30.

\bibitem[{Jiang et~al.(2024)Jiang, Liu, Ren, He, Ye, Ji, Yang, Zhang, Wei, Wang et~al.}]{jiang2024megatts2}
Jiang, Z.; Liu, J.; Ren, Y.; He, J.; Ye, Z.; Ji, S.; Yang, Q.; Zhang, C.; Wei, P.; Wang, C.; et~al. 2024.
\newblock Mega-TTS 2: Boosting Prompting Mechanisms for Zero-Shot Speech Synthesis.
\newblock In \emph{Proc. ICLR}.

\bibitem[{Kahn et~al.(2020)Kahn, Riviere, Zheng, Kharitonov, Xu, Mazar{\'e}, Karadayi, Liptchinsky, Collobert, Fuegen et~al.}]{librilight}
Kahn, J.; Riviere, M.; Zheng, W.; Kharitonov, E.; Xu, Q.; Mazar{\'e}, P.-E.; Karadayi, J.; Liptchinsky, V.; Collobert, R.; Fuegen, C.; et~al. 2020.
\newblock Libri-light: A benchmark for asr with limited or no supervision.
\newblock In \emph{Proc. ICASSP}, 7669--7673.

\bibitem[{Kong, Kim, and Bae(2020)}]{hifigan}
Kong, J.; Kim, J.; and Bae, J. 2020.
\newblock Hifi-gan: Generative adversarial networks for efficient and high fidelity speech synthesis.
\newblock \emph{Proc. NIPS}, 33: 17022--17033.

\bibitem[{Le et~al.(2024)Le, Vyas, Shi, Karrer, Sari, Moritz, Williamson, Manohar, Adi, Mahadeokar et~al.}]{le2024voicebox}
Le, M.; Vyas, A.; Shi, B.; Karrer, B.; Sari, L.; Moritz, R.; Williamson, M.; Manohar, V.; Adi, Y.; Mahadeokar, J.; et~al. 2024.
\newblock Voicebox: Text-guided multilingual universal speech generation at scale.
\newblock \emph{Proc. NIPS}, 36.

\bibitem[{Lee et~al.(2024)Lee, Kim, Kim, and Cho}]{lee2024dittotts}
Lee, K.; Kim, D.~W.; Kim, J.; and Cho, J. 2024.
\newblock DiTTo-TTS: Efficient and Scalable Zero-Shot Text-to-Speech with Diffusion Transformer.
\newblock \emph{arXiv preprint arXiv:2406.11427}.

\bibitem[{Li, Li, and Li(2023)}]{dvqvc}
Li, D.; Li, X.; and Li, X. 2023.
\newblock DVQVC: An unsupervised zero-shot voice conversion framework.
\newblock In \emph{Proc. {ICASSP}}, 1--5.

\bibitem[{Li et~al.(2024)Li, Guo, Chen, and Yu}]{sefvc}
Li, J.; Guo, Y.; Chen, X.; and Yu, K. 2024.
\newblock SEF-VC: Speaker Embedding Free Zero-Shot Voice Conversion with Cross Attention.
\newblock In \emph{Proc. {ICASSP}}, 12296--12300.

\bibitem[{Li, Han, and Mesgarani(2023{\natexlab{a}})}]{li2023slmgan}
Li, Y.~A.; Han, C.; and Mesgarani, N. 2023{\natexlab{a}}.
\newblock Slmgan: Exploiting speech language model representations for unsupervised zero-shot voice conversion in gans.
\newblock In \emph{Proc. WASPAA}, 1--5.

\bibitem[{Li, Han, and Mesgarani(2023{\natexlab{b}})}]{li2023styletts}
Li, Y.~A.; Han, C.; and Mesgarani, N. 2023{\natexlab{b}}.
\newblock Styletts-vc: One-shot voice conversion by knowledge transfer from style-based tts models.
\newblock In \emph{Proc. SLT}, 920--927.

\bibitem[{Lipman et~al.(2022)Lipman, Chen, Ben-Hamu, Nickel, and Le}]{flowmatching}
Lipman, Y.; Chen, R.~T.; Ben-Hamu, H.; Nickel, M.; and Le, M. 2022.
\newblock Flow matching for generative modeling.
\newblock \emph{arXiv preprint arXiv:2210.02747}.

\bibitem[{Liu et~al.(2023)Liu, Chen, Yuan, Mei, Liu, Mandic, Wang, and Plumbley}]{liu2023audioldm}
Liu, H.; Chen, Z.; Yuan, Y.; Mei, X.; Liu, X.; Mandic, D.; Wang, W.; and Plumbley, M.~D. 2023.
\newblock Audioldm: Text-to-audio generation with latent diffusion models.
\newblock \emph{arXiv preprint arXiv:2301.12503}.

\bibitem[{Luo and Dixon(2024)}]{timbrevc1}
Luo, Y.-J.; and Dixon, S. 2024.
\newblock Posterior Variance-Parameterised Gaussian Dropout: Improving Disentangled Sequential Autoencoders for Zero-Shot Voice Conversion.
\newblock In \emph{Proc. {ICASSP}}, 11676--11680.

\bibitem[{Mehta et~al.(2024)Mehta, Tu, Beskow, Sz{\'e}kely, and Henter}]{matchatts}
Mehta, S.; Tu, R.; Beskow, J.; Sz{\'e}kely, {\'E}.; and Henter, G.~E. 2024.
\newblock Matcha-TTS: A fast TTS architecture with conditional flow matching.
\newblock In \emph{Proc. ICASSP}, 11341--11345.

\bibitem[{M{\"u}ller(2007)}]{muller2007dynamic}
M{\"u}ller, M. 2007.
\newblock Dynamic time warping.
\newblock \emph{Information retrieval for music and motion}, 69--84.

\bibitem[{Ning et~al.(2023{\natexlab{a}})Ning, Jiang, Zhu, Yao, Wang, Xie, and Bi}]{ning2023dualvc}
Ning, Z.; Jiang, Y.; Zhu, P.; Yao, J.; Wang, S.; Xie, L.; and Bi, M. 2023{\natexlab{a}}.
\newblock Dualvc: Dual-mode voice conversion using intra-model knowledge distillation and hybrid predictive coding.
\newblock \emph{arXiv preprint arXiv:2305.12425}.

\bibitem[{Ning et~al.(2023{\natexlab{b}})Ning, Xie, Zhu, Wang, Xue, Yao, Xie, and Bi}]{ning2023expressive}
Ning, Z.; Xie, Q.; Zhu, P.; Wang, Z.; Xue, L.; Yao, J.; Xie, L.; and Bi, M. 2023{\natexlab{b}}.
\newblock Expressive-vc: Highly expressive voice conversion with attention fusion of bottleneck and perturbation features.
\newblock In \emph{Proc. ICASSP}.

\bibitem[{Peebles and Xie(2023)}]{dit}
Peebles, W.; and Xie, S. 2023.
\newblock Scalable diffusion models with transformers.
\newblock In \emph{Proc. CVPR}, 4195--4205.

\bibitem[{Perez et~al.(2018)Perez, Strub, De~Vries, Dumoulin, and Courville}]{perez2018film}
Perez, E.; Strub, F.; De~Vries, H.; Dumoulin, V.; and Courville, A. 2018.
\newblock Film: Visual reasoning with a general conditioning layer.
\newblock In \emph{Proc. AAAI}, volume~32.

\bibitem[{Popov et~al.(2021)Popov, Vovk, Gogoryan, Sadekova, Kudinov, and Wei}]{diffvc}
Popov, V.; Vovk, I.; Gogoryan, V.; Sadekova, T.; Kudinov, M.; and Wei, J. 2021.
\newblock Diffusion-based voice conversion with fast maximum likelihood sampling scheme.
\newblock \emph{arXiv preprint arXiv:2109.13821}.

\bibitem[{Qian et~al.(2019)Qian, Zhang, Chang, Yang, and Hasegawa-Johnson}]{qian2019autovc}
Qian, K.; Zhang, Y.; Chang, S.; Yang, X.; and Hasegawa-Johnson, M. 2019.
\newblock Autovc: Zero-shot voice style transfer with only autoencoder loss.
\newblock In \emph{International Conference on Machine Learning}, 5210--5219. PMLR.

\bibitem[{Radford et~al.(2019)Radford, Wu, Child, Luan, Amodei, Sutskever et~al.}]{gpt2}
Radford, A.; Wu, J.; Child, R.; Luan, D.; Amodei, D.; Sutskever, I.; et~al. 2019.
\newblock Language models are unsupervised multitask learners.
\newblock \emph{OpenAI blog}, 1(8): 9.

\bibitem[{Reddy, Gopal, and Cutler(2022)}]{reddy2022dnsmos}
Reddy, C.~K.; Gopal, V.; and Cutler, R. 2022.
\newblock DNSMOS P. 835: A non-intrusive perceptual objective speech quality metric to evaluate noise suppressors.
\newblock In \emph{Proc. ICASSP}, 886--890.

\bibitem[{Saeki et~al.(2022)Saeki, Xin, Nakata, Koriyama, Takamichi, and Saruwatari}]{utmos}
Saeki, T.; Xin, D.; Nakata, W.; Koriyama, T.; Takamichi, S.; and Saruwatari, H. 2022.
\newblock Utmos: Utokyo-sarulab system for voicemos challenge 2022.
\newblock \emph{arXiv preprint arXiv:2204.02152}.

\bibitem[{Sedgwick(2012)}]{sedgwick2012pearson}
Sedgwick, P. 2012.
\newblock Pearson’s correlation coefficient.
\newblock \emph{Bmj}, 345.

\bibitem[{Shen et~al.(2023)Shen, Ju, Tan, Liu, Leng, He, Qin, Zhao, and Bian}]{naturalspeech2}
Shen, K.; Ju, Z.; Tan, X.; Liu, Y.; Leng, Y.; He, L.; Qin, T.; Zhao, S.; and Bian, J. 2023.
\newblock Naturalspeech 2: Latent diffusion models are natural and zero-shot speech and singing synthesizers.
\newblock \emph{arXiv preprint arXiv:2304.09116}.

\bibitem[{Tong et~al.(2023)Tong, Malkin, Huguet, Zhang, Rector-Brooks, Fatras, Wolf, and Bengio}]{conditional_flowmatching}
Tong, A.; Malkin, N.; Huguet, G.; Zhang, Y.; Rector-Brooks, J.; Fatras, K.; Wolf, G.; and Bengio, Y. 2023.
\newblock Conditional flow matching: Simulation-free dynamic optimal transport.
\newblock \emph{arXiv preprint arXiv:2302.00482}, 2(3).

\bibitem[{Veaux et~al.(2019)Veaux, Yamagishi, MacDonald et~al.}]{vctk}
Veaux, C.; Yamagishi, J.; MacDonald, K.; et~al. 2019.
\newblock {CSTR VCTK} corpus: English multi-speaker corpus for cstr voice cloning toolkit.

\bibitem[{Wang et~al.(2023{\natexlab{a}})Wang, Chen, Wu, Zhang, Zhou, Liu, Chen, Liu, Wang, Li et~al.}]{valle}
Wang, C.; Chen, S.; Wu, Y.; Zhang, Z.; Zhou, L.; Liu, S.; Chen, Z.; Liu, Y.; Wang, H.; Li, J.; et~al. 2023{\natexlab{a}}.
\newblock Neural codec language models are zero-shot text to speech synthesizers.
\newblock \emph{arXiv preprint arXiv:2301.02111}.

\bibitem[{Wang et~al.(2021)Wang, Deng, Yeung, Chen, Liu, and Meng}]{wang2021vqmivc}
Wang, D.; Deng, L.; Yeung, Y.~T.; Chen, X.; Liu, X.; and Meng, H. 2021.
\newblock Vqmivc: Vector quantization and mutual information-based unsupervised speech representation disentanglement for one-shot voice conversion.
\newblock \emph{arXiv preprint arXiv:2106.10132}.

\bibitem[{Wang et~al.(2022)Wang, Zhang, Wang, Cheng, and Xiao}]{wang2022drvc}
Wang, Q.; Zhang, X.; Wang, J.; Cheng, N.; and Xiao, J. 2022.
\newblock Drvc: A framework of any-to-any voice conversion with self-supervised learning.
\newblock In \emph{Proc. ICASSP}, 3184--3188.

\bibitem[{Wang et~al.(2024)Wang, Su, Finkelstein, and Jin}]{gr0vc}
Wang, Y.; Su, J.; Finkelstein, A.; and Jin, Z. 2024.
\newblock GR0: Self-Supervised Global Representation Learning for Zero-Shot Voice Conversion.
\newblock In \emph{Proc. {ICASSP}}, 10786--10790.

\bibitem[{Wang et~al.(2023{\natexlab{b}})Wang, Chen, Xie, Tian, and Wang}]{lmvc}
Wang, Z.; Chen, Y.; Xie, L.; Tian, Q.; and Wang, Y. 2023{\natexlab{b}}.
\newblock Lm-vc: Zero-shot voice conversion via speech generation based on language models.
\newblock \emph{IEEE Signal Proces. Lett.}

\bibitem[{Yang et~al.(2024)Yang, Pan, Yao, Zhang, Ye, Zhou, Xie, Ma, and Zhao}]{yang2024takinvc}
Yang, Y.; Pan, Y.; Yao, J.; Zhang, X.; Ye, J.; Zhou, H.; Xie, L.; Ma, L.; and Zhao, J. 2024.
\newblock Takin-VC: Zero-shot Voice Conversion via Jointly Hybrid Content and Memory-Augmented Context-Aware Timbre Modeling.
\newblock \emph{arXiv preprint arXiv:2410.01350}.

\bibitem[{Yao et~al.(2023)Yao, Lei, Wang, Guo, Ning, Xie, Li, Liu, and Xie}]{yao2023preserving}
Yao, J.; Lei, Y.; Wang, Q.; Guo, P.; Ning, Z.; Xie, L.; Li, H.; Liu, J.; and Xie, D. 2023.
\newblock Preserving background sound in noise-robust voice conversion via multi-task learning.
\newblock In \emph{Proc. ICASSP}.

\bibitem[{Yao et~al.(2024)Yao, Yang, Lei, Ning, Hu, Pan, Yin, Zhou, Lu, and Xie}]{yao2024promptvc}
Yao, J.; Yang, Y.; Lei, Y.; Ning, Z.; Hu, Y.; Pan, Y.; Yin, J.; Zhou, H.; Lu, H.; and Xie, L. 2024.
\newblock Promptvc: Flexible stylistic voice conversion in latent space driven by natural language prompts.
\newblock In \emph{Proc. {ICASSP}}, 10571--10575. IEEE.

\bibitem[{Yuan et~al.(2021)Yuan, Cheng, Zhang, Hao, Gan, and Carin}]{yuan2021improving}
Yuan, S.; Cheng, P.; Zhang, R.; Hao, W.; Gan, Z.; and Carin, L. 2021.
\newblock Improving zero-shot voice style transfer via disentangled representation learning.
\newblock \emph{arXiv preprint arXiv:2103.09420}.

\bibitem[{Zhou et~al.(2022)Zhou, Sisman, Liu, and Li}]{zhou2022emotional}
Zhou, K.; Sisman, B.; Liu, R.; and Li, H. 2022.
\newblock Emotional voice conversion: Theory, databases and ESD.
\newblock \emph{Speech Communication}, 137: 1--18.

\end{thebibliography}

\end{document}